%% file: main.tex
\def\year{2019}\relax
\documentclass[letterpaper]{article} 
\usepackage{aaai19}  
\usepackage{times}  
\usepackage{helvet}  
\usepackage{courier}  
\usepackage{url}  
\usepackage{graphicx}  
\usepackage[utf8]{inputenc} 
\usepackage[T1]{fontenc}    
\usepackage{hyperref}       
\usepackage{url}            
\usepackage{booktabs}       
\usepackage{amsfonts}       
\usepackage{nicefrac}       
\usepackage{microtype}      
\usepackage{graphicx}
\usepackage{subcaption}
\usepackage[cmex10]{amsmath}
\usepackage{amssymb}
\usepackage{amsthm}
\usepackage[toc,page]{appendix}
\usepackage{algorithm}
\usepackage{algorithmic}
\usepackage{color}
\usepackage{stfloats}

\newtheorem{proposition}{Proposition}
\newtheorem{theorem}{Theorem}

\begin{document}
%
\title{Regret vs. Bandwidth Trade-off for Recommendation Systems}
\author{Linqi Song\\
City University of Hong Kong\\
linqi.song@cityu.edu.hk\\
\And
Christina Fragouli\\
University of California, Los Angeles\\
christina.fragouli@ucla.edu\\
\And
Devavrat Shah\\
Massachusetts Institute of Technology\\
devavrat@mit.edu
}
\maketitle

\begin{abstract}
We consider recommendation systems that need to operate under  wireless bandwidth constraints, measured as number of broadcast transmissions, and demonstrate a (tight for some instances) tradeoff between regret and bandwidth for two scenarios: the case of  multi-armed bandit with context, and the case where there is a latent structure in the message space that we can exploit to reduce the learning phase.
\end{abstract}

\input{intro_X.tex}
\input{System_X.tex}
\input{group_X3.tex}

\input{latent_X.tex}

\input{experiments.tex}

\newpage
\bibliographystyle{aaai}
\bibliography{rs}
\newpage
\appendix
\input{appendix_broadcastingBandits.tex}

\input{appendix_tradeoffBroadcastingRegret.tex}
\input{appendix.tex}
\input{appendix_broadcastingLatent.tex}

\end{document}

%% file: intro_X.tex
\section{Introduction}
\label{sec:intro}

With the booming of wireless devices and the pushing of computing power to the edge resources close to end users, wireless recommendation systems are  becoming increasingly popular, with applications spanning from tourism related recommendations, to mall stations serving coupons, to autonomous vehicles making recommendations to each other \cite{yang2008location} \cite{gavalas2014mobile}  \cite{woerndl2009context} \cite{ricci2010mobile}.
Works in the literature have looked at energy efficient mobile recommendation systems \cite{ge2010energy}, location aware wireless recommendation systems \cite{yang2008location}, and peer to peer connectivity in wireless recommendation systems \cite{yang2013itravel}. However, as far as we know, the existing work
has not looked into taking wireless bandwidth constraints into account. 

Bandwidth constraints can significantly affect  performance;
unsatisfactory delivery has already translated to billions in industry loss.
For instance,
viewers have low patience with poor quality video, and tend to abandon viewing within a few seconds:  it may be more profitable to recommend a video advertisement (ad) that can play seamlessly and may have a lower expected reward, than the ad that has the highest reward but cannot retain the use engagement because it has a longer start time.
The goal of this paper is to study trade-offs between  learning and wireless bandwidth.

There has been a growing  literature~\cite{li2018fundamental} \cite{song2018making} that studies distributed computing over wireless. These works capture  wireless broadcasting, i.e., the fact that over wireless when a node transmits, all nodes within the same transmission radius can overhear the transmitted messages.  We adopt this first order modeling of wireless as well  (see also \cite{Birk1998}), and measure the wireless bandwidth in terms of broadcast transmissions. However, unlike these works, our focus is on learning and recommendation systems (that cannot be abstracted within the same framework).

Our main contribution is to investigate the interactions between broadcast transmissions and learning techniques for wireless recommendation systems. We derive a trade-off relationship between the number of broadcast transmissions we utilize and the learning performance of the recommendation system in two scenarios: traditional contextual bandit framework and latent contextual bandit framework. This trade-off enables to understand what performance is possible to achieve when operating under bandwidth constraints, and is in some cases tight. Accordingly, we propose two new bandit learning frameworks:
(1) the {\em contextual broadcast bandit} framework; and
(2)  the {\em latent contextual bandit} framework. 

For the contextual broadcast bandit problem, we consider that groups of users with certain context arrival process are coming to the recommendation systems and ask how can we design the appropriate recommendation techniques. In this case, the recommendation strategy needs not only learn the preference of each individual user, but also needs to balance the learning efficiencies among users by designing broadcast schemes. We use an epoch-based scheme that explicitly splits the exploration and exploitation phases for learning and design broadcasting schemes accordingly. Through this first algorithm, we show the tradeoff behavior of this broadcast bandit: the learning speed is proportional to the square of available bandwidth (i.e., learning regret is inversely proportional to that). 

For the latent contextual bandit problem, there is an underlying clustering of messages that is unknown to the system (latent structure \cite{song2016blind}), and thus the system  needs to learn both the user preferences as well as the latent structure. 
To tackle this problem, we highlight a new {\em cross-context learning} technique, that we believe is also of independent interest: when exploring the latent structure in one context, the algorithm can utilize this knowledge in other contexts. To do this, the algorithm first needs to solve a combinatorial problem that coordinates the latent structure exploration in different contexts, and then  take advantage of this latent structure in the message space, to accelerate the learning phase. We show that, by learning the latent structure, we can realize benefits up to $O(M/L)$, where $M$ is the number of messages and $L$ is the number of message clusters. In addition, we examine how these benefits change as we enable broadcast transmissions and show that, similar to the traditional contextual bandit case, the learning speed is also proportional to the square of available bandwidth.
\vspace{-0.15in}
\paragraph*{Related Work.}
Given how ubiquitous recommendation systems are,
there is a very rich literature in the field, e.g., \cite{woerndl2009context} \cite{yang2008location}  \cite{yang2013itravel}  \cite{gavalas2014mobile} \cite{adomavicius2015context}  \cite{ricci2010mobile} ; however, these studies differ significantly from ours in the sense that they did not combine wireless broadcasting/communication features in the recommendation strategies. 

One can refer to a bulk of works for conventional multi-armed bandit problems, e.g., \cite{auer1995gambling}  \cite{auer2002finite} , and for contextual bandit problems as well, e.g., \cite{agrawal2013thompson}  \cite{slivkins2011contextual} \cite{langford2008epoch}. Closer to our work for the traditional contextual bandit problems are
works on Combinatorial Semi-Bandits (CSB) \cite{combes2015combinatorial}  \cite{gai2012combinatorial} \cite{wen2015efficient} that study the combinatorial structure between users (contexts) and messages. A main difference with our work is that, as these works do not target wireless scenarios, they do not take into account user dynamics, that translate to stochastic context arrival patterns. Moreover, the metrics we use, the trade-offs we derive and the algorithms are different. 

A recent work on latent structure and bandit framework \cite{maillard2014latent} focuses on learning which user belongs to which type given a set of types with known reward distribution; as well as tries to cluster arms into clusters (to reduce the number of arms) to facilitate the learning process. However, it does not consider  learning across contexts. In contrast, our proposed cross-context learning approach aims to extend the knowledge learned in one context to other contexts. Another stream of work studies modeling recommendation problem using latent structure \cite{song2016blind} \cite{bresler2014latent}, due to the unavailability of user or context features. 

There is also  a very rich literature on communication complexity of statistical estimation \cite{barak2013compress} \cite{amari1998statistical}, yet  such works also do not take  learning into account, and moreover the communication cost is measured in terms of the number of exchanged bits, and not in terms of wireless broadcast channel uses (these two metrics do not directly translate to each other, and in general can result in very different algorithms). In machine learning, some learning techniques, e.g., federated learning\cite{konevcny2016federated}  \cite{barcelos2011agent}, are just starting to take into account communication cost with the emergence of edge computing.  There are  some emerging studies on communication combined with learning for specific distributed tasks \cite{balcan2012distributed},  but these are not over wireless.  

Finally,  constraining the exploitation phase to use broadcast (as opposed to private) transmissions has been recently examined in the information theory literature \cite{song2018making}, where the problem was shown to be NP-hard, and approximation algorithms were proposed. 

However, as far as we know, learning with or without latent structure and broadcasting have not been examined.


%% file: System_X.tex
\section{System Model and Problem Formulation}
\label{sec:system}

\paragraph*{Traditional Contextual Bandit Framework.}
At a high level, a recommendation system has a set of messages to serve (say video ads); the system needs to learn what are the message payoffs (explore) and serve the ads that maximize the expected payoffs (exploit). The users stochastically arrive with some associated context (eg., user is in lunch break, or user is commuting), and the payoff each message achieves depends on the context.  

Recommendations based on the bandit framework deal with the { exploration and exploitation trade-off} in sequential decision making. There are many approaches to balance this trade-off, such as indexed based algorithms (e.g., UCB) \cite{auer2002finite}\cite{slivkins2011contextual}, sampling based algorithms (e.g., Thompson sampling) \cite{agrawal2013thompson}, and epoch-based algorithms \cite{langford2008epoch} \cite{vakili2013deterministic}.
As a first attempt, we will focus on  epoch-based  strategies, where the algorithm first operates  for a certain number of timeslots in an  exploration phase, learning what is the average payoff  each message can offer in a given context; and then in an  exploitation phase, where the system serves the messages it expects to have the highest payoffs. The other reason of using epoch-based strategies is for practical recommendation system deployment consideration, e.g., the wireless recommendation system operation often has some `trial period' (e.g., free of charge for the users) that can be used for the exploration phase.

We assume a finite number $M$ of messages from a message set
$\mathcal{A}=[1:M]$ and a finite number $K$ of contexts from a  set
$\mathcal{X}=[1:K]$, where $[1:X]$ denotes the set $\{1,2,\ldots,X\}$. The system operates in discrete timeslots. At each timeslot $t$:
1) A user  with some context $x_t \in \mathcal{X}$ arrives.
2) The server observes the context $x_t$, and makes a recommendation $a_t \in \mathcal{A}$, according to the algorithm it uses.
3) The server observes the payoff $r_t$ of the message $a_t$ in the context $x_t$, where $r_t = r_t(x_t, a_t)$ is a random process that depends on the message and the context.

Following the literature, we assume that the payoff $r_t(x_t,a_t)$ is generated i.i.d. according to some fixed but unknown distribution \cite{langford2008epoch}  \cite{lu2010contextual}. We denote  by $\mu(x_t,a_t)$ the expectation of $r_t(x_t,a_t)$.  Throughout the paper, we will assume that the average payoff of a message $m$ for a user with context $k$, $\mu(k,m)$, has the {\em discrete payoff structure}. Formally, this  structure assumes that  the average payoff $\mu(k,m)$ only takes discrete values\footnote{This assumption can sometimes be relaxed to be a lower bound on the payoff difference, e.g., in Alg.~1.} of $\Delta,2\Delta,3\Delta,\ldots \leq 1$. This stems from the recommendation deployment practice that many times, the ratings of a movie, a product, etc., are kept in discrete form (e.g., $3.5$ stars). We assume that the server knows a lower bound of $\Delta$, namely, $\Delta \ge \underline{\Delta}$ for some constant $\underline{\Delta}$.
We also assume that the algorithms know in advance the total number $T$ of timeslots the system needs to operate (this assumption can be relaxed to achieve the same  performance without knowing $T$ by using the ``doubling trick'' \cite{auer2002adaptive}).

The performance is  measured using the \emph{regret} $R(T)$, which is defined as the expected difference of payoffs up to time $T$ between the best possible recommendation strategy and an algorithm's recommendation strategy.  If we denote the best possible strategy as $a^*(x_t) = \arg \max_{m \in \mathcal{A}}\mu(x_t,m)$, the regret can be calculated as 
\begin{equation}
\label{eq:singleRegret}
R(T) = \mathbb{E} [\sum_{t=1}^{T} \mu(x_t, a^*(x_t)) - \sum_{t=1}^{T} r_t(x_t,a_t)]. 
\end{equation}

\paragraph*{Broadcasting in the Contextual Bandit Framework.}
We here still assume that there are $M$ messages, $K$ contexts and $T$ timeslots, however, at each timeslot $t$, a group of $N$ users arrives at the system. Each user $n\in [1:N]$ has an associated context  $x_t(n) \in [1:K]$. We denote by $x_t = (x_t(1),x_t(2),\ldots,x_t(N))$ the vector that collects the context arrivals at time $t$.

We assume that the server is wireless, and is equipped with the capability to  make lossless broadcast transmissions that deliver the same messages to multiple users \cite{Birk1998} \cite{li2018fundamental}. The server at each timeslot can make $\tau$, with $1\leq\tau\leq N$, broadcast transmissions.
One broadcast transmission  delivers the same message to all the users; $N$ transmissions deliver a private item to each of the users (for $\tau=N$ we trivially have the traditional setting). We are interested in  the trade-off between the number of broadcast transmissions the system employs in each timeslot and the regret we can achieve.

We first introduce some notation. At each timeslot,
the server observes the context vector $x_t$ and makes $\tau$  broadcast transmissions, each transmission targeted to a disjoint set of users, denoted by $\mathcal{N}_1,\ldots,\mathcal{N}_\tau$. It then observes  the payoff vector $r_t=(r_{t}(1),\ldots, r_{t}(N))$ that collects the payoffs for all $N$ users.
The payoff $r_{t}(n) = r_{t}(n,x_t(n),a_t(n))$ for each user $n \in [1:N]$ is generated i.i.d.~across  time  and users, according to some fixed but unknown distribution $\mathcal{P}_{k,m}$ that depends only on the observed context $x_t(n)=k$ and the served message for the $n$-th user $a_t(n)=m$. The expectation of $r_{t}(n)$ when observing the context  $x_t(n)$ and receiving the item  $a_t(n)$ is denoted by $\mu(x_t(n),a_t(n))$.
We will  use a vector $y_t$ of size $K$ to capture the arrival pattern, in terms of number of users in each context, for each group of $N$ users. Namely, $y_t=(y_{t,1},y_{t,2},\ldots,y_{t,K}) \in {[1:N]}^K$, where the $k$-th element represents the number of users with context $k$ arriving at time $t$, and thus $\sum_{k=1}^{K} y_{t,k} = N$. We will assume that the arrival process is i.i.d. and $y_t$ is generated according to some fixed but unknown distribution $\mathcal{P}_Y$. The \emph{arrival rate} for each context $k$ is denoted by $\nu_k$, and is the average number of users associated with context $k$ arriving at the system at each timeslot according to the distribution $\mathcal{P}_Y$. Clearly, we have $\sum_{k=1}^{K} \nu_{k} = N$. We define $\underline{\nu}$ to be the smallest arrival rate - that is, $\underline{\nu}$ is a constant that lower bounds all context arrival rates. 

Borrowing terms from online learning, we define
the {\em total regret} $L(T)=R(T)+B(T)$, to be the sum of two terms, the {\em online learning regret} $R(T)$, and the {\em broadcasting exploitation regret} $B(T)$. The learning regret $R(T)$ expresses  the average payoff difference between the best possible (offline) recommendation strategy that uses $\tau$ transmissions per timeslot, and the algorithm's (online) strategy with $\tau$ transmissions. Here, the best possible strategy for context arrival vector $x_t$ and  broadcast transmission to the users in group $\mathcal{N}_s$ is defined as
\begin{equation}
\label{eq:optimalSelection}
\begin{array}{ll}
a^*(x_t) = (a_1^*(x_t),\ldots,a_\tau^*(x_t)),
\end{array}
\end{equation}
where $a_s^*(x_t)= \arg \max_{m \in \mathcal{A}} \sum_{n \in \mathcal{N}_s} \mu(x_t(n),m)$. The best selection for the $n$-th user is denoted by $a^*(n,x_t) = a_s^*(x_t)$ if $n \in \mathcal{N}_s$. The learning regret is
\begin{equation}
\label{eq:regret}
R(T) = \mathbb{E} \sum_{t=1}^{T} \sum_{n=1}^{N} [\mu(x_t(n), a^*(n,x_t)) -  r_t(x_t(n),a_t(n))].
\end{equation}
The broadcasting exploitation regret $B(T)$ captures the performance loss using $\tau$ transmissions compared with $N$ transmissions: even if we perfectly learned what is the best message to serve to each user, we cannot do so, because we are constrained to only use $\tau$ transmissions to serve all $N$ clients. Note that $B(T)$ is not caused by online learning. 


\paragraph*{Broadcasting Exploitation Regret Bounds.}
Our work mostly focuses on designing algorithms for learning to minimize $R(T)$, that are universal (do not depend on the payoff distribution); in contrast,  $B(T)$ can be very dependent on the payoff structure, and the strategies we would use in each case can be very different. We give several examples to illustrate this in the following.

1) Diminishing $B(T)$. Clearly, if the optimal payoffs for users with different contexts concentrates on $\tau$ messages, we can find this set of $\tau$ messages and achieve $B(T) = 0$. As an extreme case, assume that the same message  is the best possible choice for every context, then we can achieve the optimal payoff with one broadcast transmission.

2) `Spike' payoff distribution. As another extreme point,  a tight worst case upper bound for $B(T)$ would be $O(TN(1-\tau/N))$.
To see this is tight, assume perfect knowledge of all distributions. Consider the simple scenario where the number of messages $M$ equals the number of contexts $K$, and is much larger than the number of users arriving at each timeslot $N$, i.e., $N << M = K$. The arrival process is uniform for users with any context, and thus the marginal arrival rate for each context is $N/K << 1$. Therefore, with high probability, the users arriving at each timeslot have different contexts. We consider a `spike' payoff distribution for messages, namely, a user with context $k$ has payoff $1$ for message $k$ and a very small payoff $\epsilon << 1$ for other messages, or formally, $\mu(k, k)=1$, and $\mu(k, k')= \epsilon$ for $k'\not= k$. Then with high probability, the payoff for each timeslot for broadcasting case is at most $\tau + (N-\tau)\epsilon \approx \tau$. However, when making $N$ transmissions at each timeslot, we get a payoff of $N$. Therefore, the performance loss is approximately $TN(1-\tau/N)$.

3) Borda score model \cite{de1781memoire}. Consider the simple scenario where the number of messages $M$ equals the number of contexts $K$, and the number of users arriving at each timeslot $N$, i.e., $\tau << N = M = K$. Each time, $N=K$ users with different contexts arrive. We consider a Latin square (scaled) Borda score payoff distribution for messages, namely, a user has payoffs $1/M,2/M,\ldots,M/M=1$ for the $M$ messages and for $K =M$ users with different contexts, every message will achieve a different payoff for different users. In other words, $\mu(k, m) \not= \mu(k', m)$ and $\mu(k, m) \not= \mu(k, m')$ for $k \not= k'$, $m\not= m'$. The payoff for each timeslot for the broadcasting case is at most $\tau(1+\frac{M-1}{M}+\ldots+\frac{M-N/\tau+1}{M})$. However, when making $N$ transmissions at each timeslot, we get a payoff of $M$. Therefore, the performance loss is approximately $OPT/(2\tau)$, where $OPT = NT$.

We can also leverage side information and  coded broadcast transmissions to increase the throughput of the system, and hence to achieve a diminishing $B(T)$, even if the optimal messages for different users are not concentrated. For example,  similar to \cite{song2018making}, in some cases, by leveraging pre-downloaded messages, even $1$ transmission can achieve optimum payoff for all users.

\paragraph*{Latent Content-Type Structure in Messages.}
In Section~\ref{sec:latent}
we will assume the following latent structure, that is fixed but unknown to the algorithms. Each message has a feature vector and according to these feature vectors, the messages are clustered into $L$ different disjoint content types, i.e., $\mathcal{A}_1, \mathcal{A}_2,\ldots,\mathcal{A}_L$, where $\sqcup_{l \in [L]}\mathcal{A}_l = \mathcal{A}$\footnote{$\sqcup$ denotes the disjoint union of sets.}. One basic observation in recommendation systems is that similar items/messages will result in similar payoffs in the same context \cite{li2010contextual} \cite{adomavicius2015context}. Motivated by this fact, the payoffs for the latent content type structure are defined as: two messages in the same content-type have the same average payoff in any context, and two messages in different content-type have different average payoffs. Formally, for two messages $a_1 \in \mathcal{A}_{l_1}$ and $a_2 \in \mathcal{A}_{l_2}$, we have $l_1=l_2$ if and only if $\mu(k,a_1) = \mu(k,a_2)$ for any context $k \in \mathcal{X}$. So that according to the discrete payoff assumption, we have $|\mu(k,a_1)- \mu(k,a_2)|\geq \Delta$ for $l_1 \not= l_2$ and any context $k \in \mathcal{X}$.
The regret with latent content-type structure is defined  as e.q.~\eqref{eq:regret}.

\paragraph*{Problem Formulations.}
In this paper, we explore:\\
$\mathbf{1.}$ What is the performance of the contextual multi-armed bandit problem when we are restricted to make one broadcast transmission per timeslot? (Section~\ref{sec:multi})\\
$\mathbf{2.}$ If we make $\tau$, with $1\leq \tau \leq N$, broadcast transmissions per timeslot, what is the achieved trade-off between regret and~$\tau$? (Section~\ref{sec:multi})\\
$\mathbf{3.}$ How do the above problems (and trade-offs) change, if we now have a latent content type structure in messages that we may also want to learn and exploit?  (Section~\ref{sec:latent})

%% file: group_X3.tex

\section{Broadcasting in the Contextual Bandits Framework}
\label{sec:multi}

We start from the case where we are restricted to make one common broadcast transmission to each group of $N$ users. We analyze the greedy  Alg.~\ref{alg:roundRobin}, that although simple, already results in an interesting trade-off curve.

\paragraph*{Algorithm 1 Description.} The algorithm does first exploration and then exploitation. For exploration, we serve each message an equal number of times, in a round robin fashion, until we have sufficiently good estimates for all contexts. For exploitation, we greedily select the message that would achieve the highest sum payoff (summing over the $N$ users) across all messages. 

We denote by $Z(k,t)$ the number of users with context $k$ arriving up to time $t$, and by $Z(k,t,m)$ the number of users the message $m$ has been recommended in context $k$ up to time $t$. We denote by $\bar r_t(k,m,Z(k,t,m))$ the sum-average payoff realization of message $m$ in context $k$ for $Z(k,t,m)$ samples up to time $t$.
Let us denote by $\mathbb{I}$ the indicator function. Formally, we have $Z(k,t) \triangleq \sum_{t' = 1}^{t} \sum_{n=1}^{N}\mathbb{I}_{\{x_{t'}(n) = k\}}$, $Z(k,t,m) \triangleq \sum_{t' = 1}^{t}  \sum_{n=1}^{N}\mathbb{I}_{\{x_{t'}(n) = k,a_{t'} = m\}}$, and
$$ \begin{array}{ll}
\bar r_t(k,m,Z(k,t,m))\triangleq \\
~~~~~~~~~\frac{1}{Z(k,t,m)} \sum_{t' = 1}^{t} \sum_{n=1}^{N} r_{t'}(k,m)\mathbb{I}_{\{x_{t'}(n) = k, a_{t'} = m\}}.
\end{array} $$
For short, we write $\bar r_t(k,m,Z(k,t,m))$ as $\bar r_t(k,m)$. 

\begin{algorithm}[tb]
   \caption{Round Robin Exploration Algorithm}
   \label{alg:roundRobin}
\begin{algorithmic}
   \STATE {\bfseries Input:} number of timeslots $T$, lower bound of discrete payoff distance $\underline{\Delta}$, and lower bound of arrival rate $\underline{\nu}$.
   \FOR{$t=1$ {\bfseries to} $T$}
   \STATE Observe the context vector $x_t$ with $N$ entries $x_t(n)$.
   \IF  [\textbf{Exploration}]{$t \leq CN^2M\log(MNKT)$}
   \STATE Recommend each message  $m \in [1:M]$ one by one over time, in a round robin fashion. 
   \ELSE [\textbf{Exploitation}]
   \STATE Recommend the message with highest sum of estimated payoffs: \\$a_t = \arg \max_{m \in [1:M]}{\sum_{n=1}^N {\bar r}_{t-1}(x_t(n),m)}$.
   \ENDIF
   \STATE Observe the payoffs of message $a_t$ for all users, $r_t(n,x_t(n),a_t)$, and update ${\bar r}_t(x_t(n),a_t)$. 
   \ENDFOR
\end{algorithmic}
\end{algorithm}

The constant parameter $C$ is defined as $C=\frac{16}{\underline{\nu}~\underline{\Delta}^2}$. Recall that $\underline{\nu}$ is a lower bound of the slowest arrival rate for all contexts (see Section~\ref{sec:system}). $\underline{\Delta}$ is a constant lower bound of the payoff gap between message selections. Note that from the discrete payoff structure and the optimal payoff defined in e.q.~\eqref{eq:optimalSelection}, we can see that the difference of the sum of payoffs for $N$ users between the optimal selection and a suboptimal selection is at least $\underline{\Delta}$, namely, either $|\sum_{n=1}^{N} [\mu(x_t(n), a^*(n,x_t)) - \mu(x_t(n), m)]| = 0$ or $|\sum_{n=1}^{N} [\mu(x_t(n), a^*(n,x_t)) - \mu(x_t(n), m)]| \geq  \underline{\Delta}$ for any context arrival $x_t$ and message $m$. Thus, it determines with what accuracy we should learn these values to be able to distinguish the optimal choice (this becomes evident in the proof of Theorem~\ref{thm1}).

\paragraph{Algorithm 1 Performance.}
\label{sec:groupRegret}
The proof of the following theorem is in Appendix~\ref{app:proofThm1}. 
\begin{theorem} \label{thm1}
The learning regret Alg. 1 achieves is  
$R(T) = O(\frac{MN^3}{\underline{\nu}~\underline{\Delta}^2}\log(MNKT)). $

\end{theorem}

\paragraph*{Regret vs. Bandwidth  Trade-off}
We here explore what are possible benefits of making multiple vs one transmission per timeslot. We will consider the case of uniform context arrivals.   Assume that we make $\tau$  transmissions  at each timeslot. We  divide the $N$ users that come to the system  into $\tau$ subgroups, where each transmission is aimed at one subgroup. As special cases, when $\tau = 1$, the problem becomes the one we discussed above in Alg.~1; when $\tau = N$, the problem becomes the traditional contextual bandit problem.

\begin{theorem}
  Given uniform context arrivals, the regret $R(T)$ is $O(\frac{MN^2K}{\tau^2} \log(MNKT))$ for $1\le \tau < N$ and  $O(MK\log(NT/K))$ for $\tau = N$.  
\end{theorem}
The proof is in Appendix~\ref{app:tradeoffBroadcastingRegret}. We see that the learning regret  reduces by a factor of $1/\tau^2$ as the number of transmissions $\tau$ increases. 
Note that for $\tau =1$ and uniform context arrival, a lower bound of regret is $\Omega(MK\log(T))$, where we require to explore each message in each context $\Omega(\log(T))$ times \cite{bubeck2012regret}. It is an open question whether our regret bound is tight in terms of the factor $N^2$, which greatly depends on the payoff distribution.

%% file: latent_X.tex
\section{Leveraging Latent Structure}
\label{sec:latent}

In this section we explore how the results change, if the messages have a latent structure, and can be clustered in content types. We start by  studying the case where there is no broadcasting (this corresponds to the point where we make $N$ transmissions in each timeslot, one for each user). We will then look at the other extreme point, where we make one broadcast transmissions per timeslot. 

\subsection{Algorithm 2: No Broadcasting}

We would  like to learn which messages belong in the same content-type; clearly, we can do so, if we find that they have the same average payoff in any one context (as per our definition in Section~\ref{sec:system} they would then have the same expected payoff in all contexts). The main question we need to answer, is how to achieve this efficiently.

\paragraph*{Illustrating Example.} We  illustrate the basic idea through a contrived example. Recall that we have $M$ messages, $K$ contexts, $L$ content-types and a fixed number of timeslots $T$ (we here assume that one user arrives per timeslot). We divide the time slots into $K$ consecutive segments, each consisting of $T/K$ slots. We consider that the context arrivals are context $1$ for the first $T/K$ time slots, context $2$ for the second $T/K$ time slots, and similarly  context $K$ for the last $T/K$ time slots. 
One straightforward approach is to solve $K$ separate multi-armed bandit sub-instances,  one for each context. In this case, for each sub-instance, each  message is served $\Theta(\log(T/K))$ times for exploration, and thus in total the regret is $O(KM\log(T/K))$.
Alternatively, we can leverage the latent structure: if the algorithm has learned  in context 1 that two messages belong in the same content type, then we have already learned how the messages are partitioned into content types, and for the remaining contexts, we just need to learn the expected payoff for one message per content type. In particular,  for the first sub-instance, we still need to select each suboptimal message $\Theta(\log(T/K))$ times. However, because for the remaining contexts we now need to explore only one message per content type, we can achieve a better regret performance $O(M\log(T/K)+(K-1)(L-1)\log(T/K)) = O((M+KL)\log(T/K))$. The regret improves by a factor of $\min\{K,M/L\}$.

\paragraph*{Algorithm 2 Description.} 
\label{sec:algorithm}
We next assume that the context arrives i.i.d. with equal probability for each context, i.e., i.i.d., with average arrival rate $\nu_1=\nu_2=\ldots=\nu_K =1/K$. However, the result can be easily extended to any fixed distribution $\nu_1,\nu_2,\ldots,\nu_K$ by scaling with a factor $\max_{k \in [1:K]}{\nu_k} /\min_{k \in [1:K]}{\nu_k}$.
Algorithm 2 uses two exploration phases that happen in parallel: one  to learn which messages are clustered in the  same content type, and the other  to learn the expected payoffs. It uses two routines; we next describe routine 1.

Routine 1 runs only once at the initialization phase to allocate ``message pairs into contexts''.
A difference from the contrived illustration example is that contexts in general arrive randomly  intertwined; and thus  it is not efficient to learn the content types within a single context, we need to partition this task among all contexts.  For each pair of messages, we may explore within a different context whether these two messages belong in the same content-type.
To do so, Routine 1 solves a set cover problem on a bipartite graph as described next.
 
\underline{Routine 1: Allocate Message Pairs in Contexts}

Create a bipartite graph, where one side lists all 
$M \choose 2$ pairs of messages as elements, and  the other side lists all $M \choose s$ distinct message subsets of size $s$  as subsets ($s$ is a parameter to be shown as $O(\frac{M\sqrt{\log(MK)}}{\sqrt{K}})$  in Appendix~\ref{app:proofCover}). We connect a subset with a pair of messages $\{m_1,m_2\}$, if and only if $m_1$ and $m_2$ are both contained in the subset. The {\em minimum set cover} is the minimum number of subsets to cover all pairs of messages (i.e, elements). The parameter $s$ is selected as the minimum integer such that the minimum set cover of the above problem\footnote{This is also referred to as the {\em covering design} problem in combinatorics.} equals $K$. Let us denote by $\mathcal{S}_1,\mathcal{S}_2,\ldots,\mathcal{S}_K$ the $K$ selected subsets, each containing $s$ messages, that cover all the message pairs. We allocate these $K$ subsets to the $K$ contexts and say the subset of messages $\mathcal{S}_k$ is associated with context $k \in [K]$.

Whenever context $k$ arrives, we will be exploring the messages in  $\mathcal{S}_k$, and thus determine whether each of the  $s \choose 2$ pairs of messages belongs in the same content type or not.\footnote{Note that Routine 1 offers just a heuristic approach in performing this allocation, and not claimed to be optimal.} Similar to Section~\ref{sec:multi}, we denote by $Z(k,t)$ the number of users with context $k$ arriving up to time $t$, and by $Z(k,t,m)$ the number of users a message $m$ has been recommended in context $k$ up to time $t$. We denote by $\bar r_t(k,m,Z(k,t,m))$, or $\bar r_t(k,m)$ for short, the sum-average payoff realization of message $m$ in context $k$ for $Z(k,t,m)$ samples up to $t$.
Then the Alg.~2 runs as follows.

\underline{Algorithm 2: Learning Latent Structure for Bandit Problem}

$\bullet$ Initialization: run Routine 1.

$\bullet$ {\bf Step}  $\mathbf{1.}$ Observe the context arrival $x_t = k \in \mathcal{A}$.

$\bullet$ {\bf Step}  $\mathbf{2.}$ Perform Exploration 1,~2, or Exploitation, as described next.

$-$ $\mathbf{2.1}$ {\em Exploration 1}: Check if  $Z(k,t,m) < D(t)$ for some $m \in \mathcal{S}_k$, with $D(t)=\frac{32}{\underline{\Delta}^2}\log(t)$, a threshold to control the number of samples of each message $m \in \mathcal{S}_k$ in each context $k$. 
That is, check if  some message in $\mathcal{S}_k$ has not been recommended $D(t)$ times. If yes, then recommend this message. If more than one are available, choose an arbitrary one. Bypass Steps 2.2-2.3 and go to Step 3. 
If no, then $Z(k,t,m) \geq D(t)$ holds for all messages $m \in \mathcal{S}_k$, and the algorithm goes to Step 2.2.

$-$ $\mathbf{2.2}$ {\em Exploration 2}: 
Cluster all the messages into content types according to Routine 2 (described later), and denote by $\mathcal{C}(t)$ the resulting set of content types. Select a message from each content type as a representative message and denote this set of messages by $\mathcal{X}^\dagger_{\mathcal{C}(t)}$. If for two different timeslots we end up with the same partition of messages into content types, we choose the same set of representatives $\mathcal{X}^\dagger_{\mathcal{C}(t)}$.
Check if some message in $\mathcal{X}^\dagger_{\mathcal{C}(t)}$ has been recommended less than $D(t)$ times in context $k$, i.e., check if $Z(k,t,m) < D(t)$ for some $m \in \mathcal{X}^\dagger_{\mathcal{C}(t)}$ (this is possible, because we clustered all messages, and not only the messages in $\mathcal{S}_k$).  If yes, then recommend this message. If more than one are available, choose an arbitrary one. Bypass Step 2.3 and go to Step 3. 
If no, then go to Step 2.3.

$-$ $\mathbf{2.3}$ {\em Exploitation}: Recommend the message in $\mathcal{X}^\dagger_{\mathcal{C}(t)}$ with the maximum sum-average payoff realization in context $k$, with ties broken arbitrarily, i.e., recommend $\arg \max_{m \in \mathcal{X}^\dagger_{\mathcal{C}(t)}} {\bar r}_{t-1}(k,m)$.

$\bullet$ {\bf Step} $\mathbf{3.}$ Let $x_t$ be the message recommended at this time slot. Observe payoff $r_t = r_t(x_t,a_t)$. 

The clustering algorithm, referred as routine 2, runs several times at Step 2, even before we have collected enough statistics to estimate the expected payoffs for each message.

\underline{Routine 2: Clustering into Content Types}\\
$\bullet$ Input: pair-wise payoff difference estimates for all pairs of messages $\{m_1,m_2\}$, $\bar d(m_1,m_2,t) \triangleq |\bar r_t(x_{m_1,m_2},m_1) - \bar r(x_{m_1,m_2},m_2)|$, where $x_{m_1,m_2}$ is the context that the pair $\{m_1,m_2\}$ is assigned to.

$\bullet$ Output: a set of content types $\mathcal{C}(t)$.

$\bullet$ Initialization: assume each message belongs in different content type and add all the resulting $M$  types in $\mathcal{C}(t)$.  

$\bullet$ Repeatedly do the following: find the minimum of $\bar{d} (\mathcal{M}_1,\mathcal{M}_2,t)$ among all content types $\mathcal{M}_1,\mathcal{M}_2 \in \mathcal{C}(t)$, where $\bar{d} (\mathcal{M}_1,\mathcal{M}_2,t)$ is calculated as the maximum estimated payoff difference between any two messages in each type, i.e., $\bar{d} (\mathcal{M}_1,\mathcal{M}_2,t) \triangleq \max_{m_1 \in \mathcal{M}_1,m_2\in \mathcal{M}_2} \bar d(m_1,m_2,t)$. For ease of notation, we may omit the variable $t$ and write as $\bar{d} (\mathcal{M}_1,\mathcal{M}_2)$. Find the minimum distance $\bar{d} (\mathcal{M}',\mathcal{M}'')$ among all pairs of content types in $\mathcal{C}(t)$. If $\bar{d} (\mathcal{M}',\mathcal{M}'') \geq \underline{\Delta}/2$, then stop and output the current set of types $\mathcal{C}(t)$; otherwise, cluster $\mathcal{M}'$ and $\mathcal{M}''$ together by removing $\mathcal{M}'$ and $\mathcal{M}''$ from $\mathcal{C}(t)$ and adding $\mathcal{M}'\cup\mathcal{M}''$ into $\mathcal{C}(t)$;  repeat the above process.

\paragraph*{Algorithm 2 Performance.} 
The proof of the following theorem is provided in Appendix~\ref{app:proofThm2}.
\begin{theorem}
\label{thm:alg2}
The learning regret of the proposed algorithm can be upper bounded by 
\begin{equation}
R(T)= O((KM\sqrt{\frac{\log(MK)}{K}}+KL)\log(T))
\end{equation}
where $L$ is the number of content types, $K$ is the number of contexts and $M$ is the number of messages.
\end{theorem}

Comparing with an algorithm that does not explore the latent structure, i.e., runs $K$ conventional multi-armed bandit algorithms with $M$ messages that achieves regret  $O(KM\log(T))$, we can gain a factor of $M/(M\sqrt{\frac{\log(MK)}{K}}+L)$ in terms of the regret performance. Note that the latent contextual bandit problem achieves at least $\Omega(KL\log(T))$ regret, which requires at least $\Omega(\log(T))$ explorations for each content type in each context. This indicates that as long as the size of the content type $M/L\le O(\sqrt{K/\log(MK)})$, then the regret achieved in Theorem~\ref{thm:alg2} is tight.

 \subsection{Algorithm 3: Broadcasting to N users}


 \paragraph*{Algorithm 3 Description.}
 Algorithm~3 builds on Algorithm 1: we add a latent structure exploration to Algorithm 1, and then recommend one message from each identified content type.

 The latent structure exploration is simpler and less efficient than what Algorithm 2 uses: we first get estimate the expected payoff for all messages under all contexts; we then cluster the messages into content types.  Thus, unlike 
Algorithm 2, we do not use Routine 1 for instance, and instead, we end up learning the expected payoffs of all messages in the contexts with maximum arrival rate that is lower bounded by $N/K$. 
 We find that with high probability, if the payoff estimates of two messages differ greater than $\underline{\Delta}/2$, then they are clustered into two content types; and that if this difference is less than $\underline{\Delta}/2$, they are clustered into the same content types. In this case, we need to estimate the payoff of a message $m$ for a context $k$ as close as $\underline{\Delta}/2$ within its true average $\mu(k,m)$.
 

\underline{Algorithm 3: Broadcasting with Latent Structure}

$\bullet$ {\em Exploration phase 1:}  If $t \le \frac{32KM}{\underline{\Delta}^2}\log(NTM)$,  performing round robin recommendation over rounds, where in each round each message is served once.

$\bullet$  {\em Clustering:} At the end of exploration phase 1, cluster the  messages using Routine 2  into $L$ content types, and select one representative message from each content type. 

$\bullet$ {\em Exploration phase 2:} If $\frac{32KM}{\underline{\Delta}^2}\log(NTM) < t \le \frac{32KM}{\underline{\Delta}^2}\log(NTM)+ \frac{16N^2L}{\underline{\nu}~\underline{\Delta}^2} \log(LNKT)$,  recommend each of the $L$ messages one by one in a round robin manner.

$\bullet$ {\em Exploitation phase: } Otherwise, perform exploitation by recommending the message among the $L$ representative messages that achieves the maximum estimated sum of payoffs.

\begin{figure*}[ht!]
  \centering
  \begin{subfigure}{.45\textwidth}
  \centering
    \includegraphics[width=1.8in]{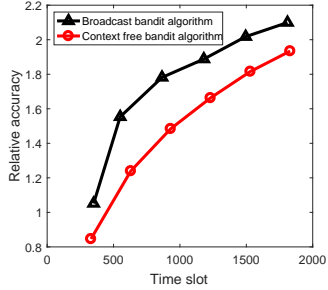}
  \caption{Broadcast bandit}
  \end{subfigure}%
  \begin{subfigure}{.45\textwidth}
  \centering
    \includegraphics[width=1.8in]{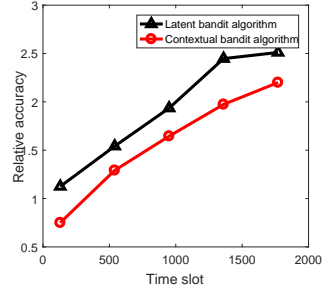}
  \caption{Latent bandit}
  \end{subfigure}
  \caption{Performance of our proposed broadcast bandit algorithm and latent bandit algorithm. Parameters in (a) are set as \# contexts $K = 5$, \# messages $M=10$, \# users per timeslot $N=10$; and in (b)  \# contexts $K = 5$, \# messages $M=50$.}
  \vspace{-0.1in}
  \label{fig:broadcastAndLatentBandit}
\end{figure*}

\paragraph*{Algorithm 3 Performance.}
The proof of the following theorem  is in Appendix~\ref{app:proofThm3}.
\begin{theorem}
\label{thm3}
The regret is bounded by
\begin{equation} \nonumber
R(T) = O(\frac{NKM}{\underline{\Delta}^2}\log(NTM)+ \frac{LN^3}{\underline{\nu}~\underline{\Delta}^2} \log(LNKT)).
\end{equation}
\end{theorem}


\paragraph*{Regret vs. Broadcasting Tradeoff.}
We explore the benefit of making multiple vs one transmission per timeslot, in terms of the achieved regret. We also consider the case of uniform context arrivals.  

$\bullet$ The regret for $\tau = 1$ and uniform context arrivals is $R(T) = O(\frac{NKM}{\underline{\Delta}^2}\log(NTM)+ \frac{LN^2K}{\underline{\Delta}^2} \log(LNKT)) = O((NKM+LN^2K)\log(NMKT))$. \\
$\bullet$ The regret for $\tau = N$ and uniform context arrivals is  $R(T) \le O((KM\sqrt{\frac{\log(MK)}{K}}+KL)\log(T))$.\\
$\bullet$ General $\tau$.  We now assume that at each timeslot, the $N$ users are divided into $\tau$ subgroups randomly, each with size $N/\tau$. This case it is equivalent to the $1$-transmission case where there are $N/\tau$ groups, $\tau T$ timeslots, $K$ contexts, and arrival rate $N/(\tau K)$ for each context. Building on the result in this Section, we get the regret $O(\frac{NKM}{\tau \underline{\Delta}^2}\log(NTM)+ \frac{LN^2K}{\tau^2 \underline{\Delta}^2} \log(LNKT))$. We see that the learning regret for different number of transmissions $\tau$ differ in a factor of $1/\tau^2$ or $1/\tau$, depending on which term dominates.

\begin{proposition}
Given uniform context arrivals, the regret $R(T)$ of latent contextual bandit is $O((NKM+LN^2K)\log(NMKT))$ for $\tau = 1$ , is $O((KM\sqrt{\frac{\log(MK)}{K}}+KL)\log(T))$ for $\tau = N$, and is  $O(\frac{NKM}{\tau \underline{\Delta}^2}\log(NTM)+ \frac{LN^2K}{\tau^2 \underline{\Delta}^2} \log(LNKT))$ for general $1\le \tau \le N$.
\end{proposition}


%% file: experiments.tex
\section{Experiments}

We conduct experiments over the Yahoo! Webscope dataset R6A, Yahoo Today Module data set\footnote{https://webscope.sandbox.yahoo.com}. This dataset is suitable for news recommendation by exploring contextual information. This dataset contains around 43 million user-news interactions (click or not) collected from Yahoo! Front Page during May 1 to 10, 2009. There are 271 news articles to be recommended. Each instance of the dataset includes the recommended news IDs and its features, the user context, and the interaction between user and news (whether the user clicked the news or not). Each context is a 5 dimensional vector, which is obtained from a higher dimensional space, that describes the user's features \cite{li2010contextual}, such as demographic information (gender and age),  geographic features, behavioral categories (about 1000 binary categories that summarize the user's consumption history). The special data collection method \cite{li2010contextual} allows evaluating online algorithms on this dataset without introducing bias.

The performance is evaluated as the relative accuracy (click through rate of applied algorithm over random selection). The experiment parameters are shown as in the caption of Fig.~\ref{fig:broadcastAndLatentBandit}. We first show the broadcast bandit performance in Fig.~\ref{fig:broadcastAndLatentBandit} (a). 
We compare our broadcast bandit algorithm with the context free algorithm, which does not take into account the contexts of users and treats the $N$ users arriving at each time slot equally to run a $M = 10$ traditional multi-armed bandit algorithm. As shown in the figure, the broadcast bandit algorithm achieves around $7\%$ performance gains. We next show the performance of our proposed latent contextual bandit algorithm in Fig.~\ref{fig:broadcastAndLatentBandit} (b), by comparing with the traditional contextual bandit algorithm that does not explore the latent structure. We show that the by exploring the latent structure, our proposed algorithm achieves around $12\%$ performance gains.

%% file: appendix_broadcastingBandits.tex
\section{Proof of Theorem~\ref{thm1}}
\label{app:proofThm1}  
We can bound the regret as 
\begin{equation}
\begin{array}{ll}
R(T) \leq R_1(T) + R_2(T) + R_3(T),
\end{array}
\notag
\end{equation} 
where $R_1(T)$ is caused by abnormal context arrival; $R_2(T)$ is caused by loss due to exploration; and $R_3(T)$ is caused by abnormal payoff realization. Next, we will show that $R_1(T)$ and $R_3(T)$ are bounded by one, and thus $R(T)$ is equal to the order of $R_2(T)$. 


We will calculate the probability that a message $m$ is recommended to less than $\frac{8N^2}{\underline{\Delta}^2}\log(MNKT)$ users associated with context $k$ during the exploration phase.  Define $T_1=\frac{16MN^2}{\underline{\nu}~\underline{\Delta}^2}\log(MNKT)$ as the number of time slots for exploration. Since the round-robin exploration phase lasts for  $T_1$ timeslots, each  message $m$ is selected $T_1/M$ times.   Let us denote by $Z(k,T_1,m)$ the total number of users with context $k$ that receive message $m$ for the exploration phase.

 We have $\mathbb{E}Z(k,T_1,m) = \frac{16N^2\nu_k}{\underline{\nu}~\underline{\Delta}^2}\log(MNKT) \geq \frac{16N^2}{\underline{\Delta}^2}\log(MNKT)$, and thus 
\begin{equation}
\begin{array}{ll}
\Pr\{\text{Message $m$ is recommended to less than} \\
\text{~~~~~~~$\frac{8N^2}{\underline{\Delta}^2}\log(MNKT)$ users with context $k$}\} \\
\leq \Pr\{Z(k,T_1,m)/N \le 0.5 \mathbb{E}Z(k,T_1,m)/N\} \\
\leq \exp(-\frac{\mathbb{E}Z(k,T_1,m)}{8N}) \\
\leq \frac{1}{(MNKT)^2}\exp(-\frac{N}{\underline{\Delta}^2}), 
\end{array}
\end{equation}
where the second inequality is from the Chernoff bound.

We say that we observe a normal context arrival, if, during the exploration phase, any message $m$, $\forall m \in[1:M]$, is recommended to more than $\frac{8N^2}{\underline{\Delta}^2}\log(MNKT)$ users with any context $k$, $\forall k \in [1:K]$, and abnormal otherwise.
Using the union bound, we can lower bound the probability of normal context arrival as follows.
\begin{equation}
\begin{array}{ll}
\Pr\{\text{Normal context arrival}\} \\
\geq 1-  \frac{MK}{(MNKT)^2}\exp(-\frac{N}{\underline{\Delta}^2}) \\
\geq 1- \frac{1}{MN^2KT^2}.
\end{array}
\end{equation}

If normal context arrival occurs, the sum-average estimate $\bar r_t(k,m)$ with at least $\frac{8N^2}{\underline{\Delta}^2}\log(MNKT)$ samples diverges from its expected value $\mu(k,m)$ with probability
\begin{equation}
\begin{array}{ll}
\Pr\{|\bar r_t(k,m) - \mu(k,m)| \geq \frac{\underline{\Delta}}{2N}\} \\
\leq 2\exp(-2\frac{8N^2}{\underline{\Delta}^2}\log(MNKT)(\frac{\underline{\Delta}}{2N})^2)\\
= \frac{2}{(MNKT)^4}.
\end{array}
\end{equation}
Accordingly, we say that we have a  normal payoff realization if the event $\{|\bar r_t(k,m) - \mu(k,m)| \leq \frac{\underline{\Delta}}{2N}\}$.

We note that if normal payoff realization occurs for all contexts and messages, then we will recommend the optimal item during the exploitation phase. Indeed, for any suboptimal message $m$, we have
\begin{equation}
  \begin{aligned}
    &\sum_{n=1}^{N}\bar{r}_t(x_t(n),a_t^*) \geq \sum_{n=1}^{N}\mu(x_t(n),a_t^*) - \frac{\underline{\Delta}}{2} \\
    &\geq \sum_{n=1}^{N}\mu(x_t(n),m) + \frac{\underline{\Delta}}{2} \geq \sum_{n=1}^{N}\bar{r}_t(x_t(n),m).
  \end{aligned}
  \end{equation}

Therefore, it holds that
\begin{equation}
\begin{array}{ll}
R_1(T) \leq \frac{NT}{MN^2KT^2} \leq 1,
\end{array}
\end{equation} 
\begin{equation}
\begin{array}{ll}
R_2(T) \leq NT_1 \leq O(\frac{MN^3}{\underline{\nu}~\underline{\Delta}^2}\log(MNKT)),
\end{array}
\end{equation} 
\begin{equation}
\begin{array}{ll}
R_3(T) \leq \frac{2MKNT}{(MNKT)^4} \leq 1,
\end{array}
\end{equation} 
which concludes the proof.

%% file: appendix_tradeoffBroadcastingRegret.tex
\section{Tradeoff Between Regret and Broadcasting}
\label{app:tradeoffBroadcastingRegret}

$\bullet$ The regret for $\tau = 1$ and uniform context arrivals is $O(MN^2K\log(MNKT))+OPT(1-\frac{1}{N})$. This is obtained from Section~\ref{sec:groupRegret}, by substituting $\underline{\nu} = N/K$ and ignoring the $1/\underline{\Delta}^2$ terms which is a constant.

$\bullet$ The regret for $\tau = N$ and uniform context arrivals is  $O(MK\log(NT/K))$. This can be derived directly from the literature \cite{auer2002finite} \cite{auer1995gambling} \cite{lu2010contextual}. There are in total $NT$ users arriving at the system and for each context, so that the average number of users arriving in each context is $NT/K$; as a result, the regret for each context is $O(M\log(NT/K))$ and the total regret is $O(MK\log(NT/K))$.  

$\bullet$ General $\tau$.  We now assume that at each timeslot, the $N$ users are divided into $\tau$ subgroups randomly, each with size $N/\tau$. This case  is equivalent to the $1$-transmission case where there are $N/\tau$ groups, $\tau T$ timeslots, $K$ contexts, and arrival rate $N/(\tau K)$ for each context. Building on the result in Section~\ref{sec:groupRegret}, we get the regret $O(\frac{MN^2K}{\tau^2} \log(MNKT))$.

\begin{figure}
\centering
\includegraphics[width=2.6in]{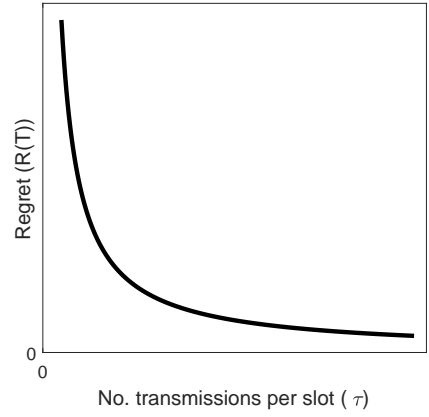}
\caption{Trade-off between the number of transmissions per timeslot $\tau$ and the regret $R(T)$ for group transmissions up to $T$.}
\vspace{-0.1in}
\label{fig:tradeOff}
\end{figure}

We see that the learning regret for different number of transmissions $\tau$ differ in a factor of $1/\tau^2$. Fig.~\ref{fig:tradeOff} plots this  versus the number of transmissions $\tau$ per timeslot.  Note that as the number of transmission increases, the learning regret decreases and this implies that the algorithm learns faster for larger $\tau$, as expected.

%% file: appendix.tex
\section{Determining the Size of Subsets}
\label{app:proofCover}
In this appendix, we will determine the size of subsets for the set covering problem. Let us first reiterate the problem, and then give a solution for the problem.

\paragraph{Problem Description.} Given a set of $M$ elements (simply $[1:M]$), a $s$-element subset is defined as a subset of $[1:M]$ that contains exactly $s$ elements. The covering design is to find smallest possible size $K^\dagger$ of a collection $\{\mathcal{S}_1,\mathcal{S}_2,\ldots,\mathcal{S}_{K^\dagger}\}$ of $s$-element subsets such that any pair of elements (any $2$-element subset) is contained in at least one selected $s$-element subset.  Now we would like to find the minimum $s$ such that the size $K^\dagger \leq K$.

\paragraph{Result.} We use a probabilistic argument to show the result that $s = O(\frac{M\sqrt{\log(MK)}}{\sqrt{K}})$ is enough for a $K$ subsets to cover all the pairs. 

We assume that we uniformly at random choose $K$ $s$-element subsets. Each pair, say $\{m_1,m_2\}$, is contained in $M-2 \choose s-2$ $s$-element subsets. So that for the randomly selected $K$ $s$-element subsets, a pair is contained in one of them with probability 
\begin{equation}
\begin{array}{ll}
\Pr\{\text{$\{m_1,m_2\}$ is not contained in any selected subset}\} \\
\overset{(a)}{=} \displaystyle \frac{{ {M \choose s} - {M-2 \choose s-2} \choose K }}{ {{M \choose s} \choose K} } 
\overset{(b)}{\leq} \displaystyle \frac{\frac{\left [{M \choose s} - {M-2 \choose s-2} \right ]^K}{K!}}{\frac{{M \choose s}^K}{4K!}} \\
\leq 4\left [1-\frac{{M-2 \choose s-2}}{{M \choose s}} \right ]^K 
\leq 4\left [1-\frac{(s-1)^2}{M^2} \right ]^K \\
\overset{(c)}{\leq} 4e^{-2\log(MK)} = \frac{4}{M^2K^2},
\end{array}
\end{equation}
where $(b)$ comes from the inequalities $\frac{n^k}{4k!}\leq {n \choose k} \leq \frac{n^k}{k!}$ for $n \geq k^2$; $(c)$ follows from the inequality $(1+\frac{x}{n})^n \leq e^x$ and the assignment $s = \frac{M\sqrt{2\log(MK)}}{\sqrt{K}}+ 1$; and $(a)$ comes from the following. There are in total ${M \choose s}$ $s$-element subsets. The denominator of the first equality $(a)$ represents the total number of ways to select $K$ subsets among all ${M \choose s}$ $s$-element subsets. There are in total ${M-2 \choose s-2}$ $s$-element subsets that contain a given pair $\{m_1,m_2\}$.  The nominator of the first equality $(a)$ represents the number of possibilities for the $K$ subsets that does not contain a given pair $\{m_1,m_2\}$. Since the $K$ subsets are selected uniformly at random, the equality $(a)$ then follows.

Given a random selection of $K$ subsets, among all ${M \choose 2}$ pairs, the average number of pairs that are not contained in any of the $K$ subsets is 
\begin{equation}
\begin{array}{ll}
{M \choose 2} \Pr\{\text{$\{m_1,m_2\}$ is not contained in any selected subset}\} \\
\leq \frac{4M^2}{M^2K^2} \leq 1.
\end{array}
\end{equation}
This implies that among all selections of the $K$ subsets, there must exists one that can cover all the pairs. Thus, we get $s = O(\frac{M\sqrt{\log(MK)}}{\sqrt{K}})$.

\section{Proof of Theorem~\ref{thm:alg2}}
\label{app:proofThm2}

We denote by $E_{t,1}$, $E_{t,2}$, and $E_{t,3}$ the events that the timeslot $t$ is an exploration 1 phase, an exploration 2 phase, and an exploitation phase, respectively. Let us denote by $\mathcal{X}^*(a_t)$ the optimal content type corresponds to the context $a_t$. Then, we can bound the regret as follows:
\begin{equation}
R(T) \le \sum_{t=1}^{T} \mathbb{E} [\mathbb{I}_{\{E_{t,1}\}} + \mathbb{I}_{\{E_{t,2}\}} + \mathbb{I}_{\{E_{t,3},x_t \notin \mathcal{X}^*(a_t)\}}], 
\end{equation}
where the first two terms in the expectation are caused by exploration, and the third term is caused by suboptimal message selection in the exploitation phase. For the second and third terms, we need to consider two scenarios: the clustering process outputs the correct content types and it does not. For simplicity, we denote by $\mathcal{C}^*$ the correct clustering $\{\mathcal{X}_1, \mathcal{X}_2,\ldots,\mathcal{X}_L\}$ and by $\mathcal{C}^*_{-}$ otherwise. By a little abuse of notation, we denote by $\mathcal{C}^*(t)$ the event that the clustering outputs $\mathcal{C}^*$ at timeslot $t$ and by $\mathcal{C}^*_{-}(t)$ the clustering process outputs a different set of content types. Therefore, the above regret can be split into $4$ terms: $R(T) \le R_1(T) + R_2(T) +R_3(T) + R_4(T)$, where 

\begin{equation}
R_1(T) = \sum_{t=1}^{T} \mathbb{E} [\mathbb{I}_{\{E_{t,1}\}}],
\end{equation}
\begin{equation}
\begin{array}{ll}
R_2(T) & = \sum_{t=1}^{T} \mathbb{E} [\mathbb{I}_{\{E_{t,2} \lor E_{t,3},\mathcal{C}^*_{-}(t)\}}] \\
& = \sum_{t=1}^{T} \Pr\{E_{t,2} \lor E_{t,3},\mathcal{C}^*_{-}(t)\} \\
& = \sum_{t=1}^{T} \Pr\{E_{t,2} \lor E_{t,3}\} \Pr\{\mathcal{C}^*_{-}(t)|E_{t,2} \lor E_{t,3}\} \\
& \le \sum_{t=1}^{T} \Pr\{\mathcal{C}^*_{-}(t)|E_{t,2} \lor E_{t,3}\}
\end{array}
\end{equation}
\begin{equation}
R_3(T) = \sum_{t=1}^{T} \mathbb{E} [\mathbb{I}_{\{E_{t,2},\mathcal{C}^*(t)\}}],
\end{equation}
and
\begin{equation}
\begin{array}{ll}
R_4(T) &= \sum_{t=1}^{T} \mathbb{E} [\mathbb{I}_{\{E_{t,3},x_t \notin \mathcal{X}^*(a_t),\mathcal{C}^*(t)\}}] \\
& = \sum_{t=1}^{T} \Pr\{E_{t,3},x_t \notin \mathcal{X}^*(a_t),\mathcal{C}^*(t)\} \\
& \le \sum_{t=1}^{T} \Pr\{x_t \notin \mathcal{X}^*(a_t)|E_{t,3},\mathcal{C}^*(t)\}
\end{array}
\end{equation}

Then $R_1(T)$ is caused by the exploration 1 and can be bounded by $sKD(T)$ based on our proposed learning algorithm, since the algorithm performs the first type exploration at most $D(T)$ times in each of $K$ contexts and the number of messages to explore in each context is at most $s$.

The second regret term $R_2(T)$ is caused by mis-clustering and we will show that the probability of mis-clustering $\Pr\{\mathcal{C}^*_{-}(t)|E_{t,2} \lor E_{t,3}\}$ is exponentially small. 

The third regret term $R_3(T)$ is caused by exploration 2 when the clustering is correct, and $R_3(T)$ can be bounded by $KLD(T)$, since when the clustering is correct, the algorithm performs the exploration 2 phases for each representative message in $\mathcal{X}^\dagger_{\mathcal{C}(t)}$ at most $D(T)$ times in each of $K$ contexts and the number of representative messages to explore in each context is $L$. 

The fourth regret term $R_4(T)$ is caused by suboptimal message selection in the exploitation phase when the clustering is correct. We will show that the probability of choosing a suboptimal message in this case is exponentially small.

In the following, we will focus on the calculations of two probabilities $\Pr\{\mathcal{C}^*_{-}(t)|E_{t,2} \lor E_{t,3}\}$ and $\Pr\{x_t \notin \mathcal{X}^*(a_t)|E_{t,3},\mathcal{C}^*(t)\}$.

We next show how to bound the mis-clustering probability if the clustering algorithm is carried out at current timeslot $t$, i.e., not an exploration 1 phase. To bound this, we define the following {\em normal} and {\em abnormal} events.

$\bullet$ {\em Context arrival abnormality}. We define the event $\{Z(k,t) \le \frac{t}{2K}, \text{ or }Z(k,t) \ge \frac{3t}{2K} \} \triangleq W_{k,t}$ as the abnormal context arrival for context $k$ at timeslot $t$. In contrast, we define $W^C_{k,t}$ as the normal context arrival for context $k$ at timeslot $t$. We denote by $W_t$ the event that there exists at least one context that has the abnormal arrival at timeslot $t$, i.e., $W_t = \lor_{k \in [1:K]} W_{k,t}$. Similarly, we define $W^C_t$ as the normal context arrival for all contexts at time $t$, i.e., the complement of $W_t$.

$\bullet$ {\em Payoff realization abnormality}. We define the event $\{|\bar r_t(k,m) - \mu(k,m)| \ge \sqrt{\frac{\log (t)}{Z(k,t,m)}} \triangleq \delta(k,t,m)\} \triangleq V_{k,m,t}$ as the abnormal payoff realization for message $m$ in context $k$ at time $t$. We denote by $V_t$ the event that there exists at least one context and one message that has abnormal payoff realization, i.e., $V_t = \lor_{k \in [1:K],m \in [1:M]} V_{k,m,t}$. Similarly, we define $V^C_t$ as the normal payoff realization for all contexts all messages at time $t$, i.e., the complement of $V_t$.

We will use the Chernoff-Hoeffding inequality later, so we recall the inequality as follows. Given $Y_i \in [0,1]$ and $Y = \sum^n_{i = 1} Y_i /n$, then the following inequality holds
\begin{equation}
\Pr\{|Y - \mathbb{E}Y| \ge u\} \le 2 e^{-2nu^2}.
\end{equation}

Next, we calculate the probability of context arrival abnormality. It is not hard to see that $\mathbb{E}[Z(k,t)] = t/K$. Using Chernoff-Hoeffding inequality, we have the probability of abnormal context arrival for context $k$ at timeslot $t$:
\begin{equation}
\begin{array}{ll}
\Pr\{W_{k,t}\} = \Pr\{|Z(k,t) - \mathbb{E}[Z(k,t)]| \geq \mathbb{E}[Z(k,t)]/2\} \\
\le 2 e^{-2 \frac{t^2}{4tK^2}} = 2e^{-\frac{t}{2K^2}}.
\end{array}
\end{equation}

If $K^2 \le \frac{t}{8 \log(t)}$, we then have
\begin{equation}
\Pr\{W_{k,t}\} \le 2e^{-4\log t} = \frac{2}{t^4},
\end{equation}
and the probability of abnormal context arrival for all contexts can be bounded by
\begin{equation}
\Pr\{W_{t}\} \le \frac{2K}{t^4} \le \frac{1}{t^2}.
\end{equation}

Next, we show that when a normal context arrival occurs for all contexts and the clustering is being processed at timeslot $t$ with $Ks \le \frac{t}{D(t)}$, the number of times a message $m \in \mathcal{S}_{k'}$ for every context $k' \in [1:K]$ is recommended at least $\frac{D(t)}{2}$ times, i.e., $Z(k',t,m) \ge \frac{D(t)}{2}$ for any $k' \in [1:K]$ and $m \in \mathcal{S}_{k'}$.

Indeed, if we consider any context $k'$ at timeslot $t$, the number of context arrivals $Z(k',t)$ is between $\frac{t}{2K}$ and $\frac{3t}{2K}$ if $W^C_t$ is true. We set $t' = t/6$ and then the number of context arrivals $Z(k',t')$ is between $\frac{t}{12K}$ and $\frac{t}{4K}$ if $W^C_{t'}$ is true. Then, we have the number of context arrivals between $t'$ and $t$ in context $k'$ is at least $Z(k',t) - Z(k',t') \geq \frac{t}{2K} -\frac{t}{4K} = \frac{t}{4K}$. However, according to our algorithm, at timeslot $t'' \ge t'$, if any message in $\mathcal{S}_{k'}$ is not recommended $D(t'') \ge D(t')$ times, then this message is to be recommended by the algorithm. Since the exploration 1 phases in context $k'$ up to time $t$ is at most $sD(t)  < \frac{t}{4K}\le Z(k',t)-Z(k',t')$, for timeslots corresponding to any of the context $k'$ arrivals between $t'$ and $t$, we will need to recommend a message in $\mathcal{S}_{k'}$ to do the exploration 1 resulting in total at least $D(t')$ times. Hence, $Z(k',t,m) \ge D(t/6) \ge \frac{D(t)}{2}$ if $W^C_t$ and $W^C_{t/6}$ are both true.

We then calculate $\Pr\{V^C_t\}$. Using the Chernoff-Hoeffding inequality, we first  bound
\begin{equation}
\begin{array}{ll}
\Pr\{V_{k,t,m}\} &= \Pr \{|\bar r_t(k,m) - \mu(k,m)| \ge \sqrt{\frac{\log (t)}{Z(k,t,m)}}\} \\
& \le 2e^{-2Z(k,t,m)\frac{\log(t)}{Z(k,t,m)}} \\
& =\frac{2}{t^2}.
\end{array}
\end{equation}

Therefore, we have
\begin{equation}
\begin{array}{ll}
\Pr\{V_t\} \le KM \Pr\{V_{k,t,m}\} \le \frac{2MK}{t^2}.
\end{array}
\end{equation}

We make the following claim to show our result for $R_2(T)$: if $W^C_t$, $W^C_{t/6}$, and $V^C_t$ hold true, then the clustering process at timeslot $t$ outputs the correct clustering $\mathcal{C}(t) = \mathcal{C}^*$.

Since if $W^C_t$ and $W^C_{t/6}$ hold true, for any $k' \in [1:K]$, $m \in \mathcal{S}_{k'}$, we have $Z(k',t,m) \ge \frac{1}{2}D(t)$ as shown above. If $V^C_t$ also holds true, we can bound $|\bar r_t(k',m) - \mu(k',m)| \leq \sqrt{\frac{\log (t)}{Z(k',t,m)}} \triangleq \delta(k',t,m)$ by 
\begin{equation}
\delta(k',t,m) = \sqrt{\frac{\log (t)}{Z(k',t,m)}} \le \sqrt{\underline{\Delta}^2/16} = \underline{\Delta}/{4},
\end{equation}
for any $k' \in [1:K]$, $m \in \mathcal{S}_{k'}$.

Recall that $\underline{\Delta}/{2}$ is the parameter to control the clustering algorithm, such that any two messages in a content type has difference of estimated payoff no more than $\underline{\Delta}/{2}$. 

For the output $\mathcal{C}(t)$, for simplicity, we denote by $\bar r_1,\mu_1$ and $\bar r_2,\mu_2$ the sum-average realized payoffs and the ground truth expected values for two messages $m_1$ and $m_2$. If the two messages $m_1$ and $m_2$ are clustered in the same content type, then we have 
\begin{equation}
|\bar r_1 -\mu_1| < \underline{\Delta}/{4},~|\bar r_2 - \mu_2| < \underline{\Delta}/{4},~|\bar r_1 - \bar r_2| \le \underline{\Delta}/{2},
\end{equation}
where the first two inequalities hold from the normal events conditions and the third one holds according to the clustering algorithm.
Therefore, we have 
\begin{equation}
\label{eq:inAcluster}
\begin{array}{ll}
|\mu_1 -\mu_2| &= |(\mu_1 - \bar r_1) - (\mu_2 -\bar r_2) + (\bar r_1 - \bar r_2)|  \\
&< |\mu_1 - \bar r_1| + |\mu_2 -\bar r_2| + |\bar r_1 - \bar r_2| \\
& \le \underline{\Delta}/{4} + \underline{\Delta}/{4} + \underline{\Delta}/{2} = \underline{\Delta}.
\end{array}
\end{equation}

Since $\underline{\Delta}$ is no more than the minimum gap between two different types, we have $\mu_1 = \mu_2$.

If two messages $m_1 \in \mathcal{M}_1$ and $m_2 \in \mathcal{M}_2$ are clustered into two different content types $\mathcal{M}_1$ and $\mathcal{M}_2$. Then, we have  
\begin{equation}
|\bar r_1 -\mu_1| < \underline{\Delta}/{4},|\bar r_2 - \mu_2| <  \underline{\Delta}/{4},\bar d(\mathcal{M}_1,\mathcal{M}_1) >  \underline{\Delta}/{2},
\end{equation}
where the first two inequalities hold from the normal events conditions, and the third one holds according to the clustering algorithm. Recall that $\bar d(\mathcal{M}_1,\mathcal{M}_2) = \max_{m'_1 \in \mathcal{M}_1,m'_2 \in \mathcal{M}_2} |\bar d(m'_1,m'_2)|$, 
so there exist some $m^*_1 \in \mathcal{M}_1$ and $m^*_2 \in \mathcal{M}_2$ such that $|\bar r_{m^*_1} -\bar r_{m^*_2}| > \underline{\Delta}/{2}$.
If $|\bar r_1 -\bar r_2| > \underline{\Delta}/{2}$, we have
\begin{equation}
\label{eq:betweenClusters}
\begin{array}{ll}
|\mu_1 - \mu_2| &= |(\mu_1 - \bar r_1) - (\mu_2 -\bar r_2) + (\bar r_1 - \bar r_2)|  \\
& \ge |\bar r_1 - \bar r_2| - |\mu_1 - \bar r_1| - |\mu_2 -\bar r_2|  \\
& > \underline{\Delta}/{2} -\underline{\Delta}/{4} -\underline{\Delta}/{4} = 0,
\end{array}
\end{equation}
which implies that $\mu_1 \not= \mu_2$.

If $|\bar r_1 -\bar r_2| \le \underline{\Delta}/{2}$, then from the above proof in eq. \eqref{eq:inAcluster}, we can see that $\mu_1 = \mu_2$. We also have $\mu_1 = \mu_{m^*_1}$ and $\mu_2 = \mu_{m^*_2}$, and hence $\mu_{m^*_1} = \mu_{m^*_2} $. However, since $|\bar r_{m^*_1} -\bar r_{m^*_2}| > \underline{\Delta}/{2}$, we have $\mu_{m^*_1} \not= \mu_{m^*_2} $ from eq.~\eqref{eq:betweenClusters}, resulting in a contradiction.

Therefore, we can bound the probability of mis-clustering by
\begin{equation}
\begin{array}{ll}
\Pr\{\mathcal{C}^*_{-}(t)|E_{t,2} \lor E_{t,3}\} \le \Pr\{W_t\} + \Pr\{W_{t/6}\} + \Pr\{V_t\} \\
\le \frac{1}{t^2} + \frac{36}{t^2} + \frac{2MK}{t^2}.
\end{array}
\end{equation}
The regret $R_2(T)$ can then be bounded by
\begin{equation}
R_2(T) \le (2MK+37)\pi^2/6,
\end{equation}
where we use the equality $\sum_{t=1}^{\infty} 1/t^2 = \pi^2 /6$.

In the following, we calculate the regret $R_4(T)$. We argue that given the correct clustering $\mathcal{C}^*(t)$, if $V^C_t$ holds true, then the suboptimal message cannot be selected at timeslot $t$ in an exploitation phase.

Indeed, let us consider the set of representative messages corresponding to the correct clustering $\mathcal{C}^*$, denoted by $\mathcal{X}^{\dagger*}$. For simplicity, let us denote by $\bar r^*,\mu^*$ and $\bar r_1,\mu_1$ the sum-average estimates and the expected value of the payoffs for the best message $m^* = x^*(a_t) \in \mathcal{X}^{\dagger*}$ and any suboptimal message $m_1 \in \mathcal{X}^{\dagger*}$. We have $|\mu^* - \mu_1| \ge \Delta > \underline{\Delta}$, $|\mu^* -\bar r^*| < \underline{\Delta}/4$, $|\mu_1 -\bar r_1| < \underline{\Delta}/4$, where the first inequality comes from the definition of $\underline{\Delta}$, the last two inequalities follow from that for a normal event, the number of recommendations for each message in $\mathcal{X}^{\dagger*}$ is at least $D(t)$ so as to trigger an exploitation phase. Therefore,
\begin{equation}
\bar r^* - \bar r_1 > \mu^* - \underline{\Delta}/4 - (\mu_1 + \underline{\Delta}/4) \ge \underline{\Delta}/2 >0.
\end{equation}
This implies that the probability that a suboptimal message is selected in the exploitation phase can be bounded by
\begin{equation}
\Pr\{x_t \notin \mathcal{X}^*(a_t)|E_{t,3},\mathcal{C}^*(t)\} \le \Pr\{V_t\} \le \frac{2MK}{t^2}.
\end{equation}
Then the regret $R_4(T)$ can be bounded by $MK\pi^2/3$.

To sum up, we get the regret $R(T) \le K(s+L)D(T)+(4MK+37)\pi^2/6 = O(\frac{K(s+L)}{\underline{\Delta}^2}\log(T))$. Plugging in $s$ from Appendix~\ref{app:proofCover}, we get the result.

%% file: appendix_broadcastingLatent.tex
\section{Proof of Theorem~\ref{thm3}}
\label{app:proofThm3}
First note that in the exploration phase 1, the round robin recommendation is performed at least for $\frac{32K}{\underline{\Delta}^2}\log(KTM)$ rounds (in every round each of the $M$ messages is recommended once).

We can then bound the probability that a pair of messages $\{m_1,m_2\}$ is sampled more than $\frac{16}{\underline{\Delta}^2}\log(NTM)$ times in some context $k$. Let us denote by $Z(k,m_1)$ the number of users with context $k$ that are recommended a message $m_1$ in this phase. Let us choose the context $k$ with the maximum arrival rate, then clearly $\nu_k \geq N/K$. Then $\mathbb{E} Z(k,m_1) \ge \frac{32K \nu_k }{\underline{\Delta}^2}\log(NTM) \geq  \frac{32N}{\underline{\Delta}^2}\log(NTM)$.

\begin{equation}\nonumber
\begin{array}{ll}
\Pr\{\text{a pair of messages $\{m_1,m_2\}$ is recommended to no} \\
\text{~~~~ more than $\frac{16}{\underline{\Delta}^2}\log(NTM)$ users in context $k$} \} \\
\leq 2 \Pr\{\text{message $m_1$ is recommended no more than} \\
\text{~~~~~ $\frac{16}{\underline{\Delta}^2}\log(NTM)$ times in some context $k$} \} \\
\leq 2 \Pr\{\frac{Z(k,m_1)}{N} \leq \frac{\mathbb{E}Z(k,m_1)}{2N}\} \\
\leq 2 \exp(-\frac{32N}{8N\underline{\Delta}^2} \log(NMT)) \\
\leq \frac{1}{(NMT)^4}.
\end{array}
\end{equation}
Using the union bound, the probability that all pairs of messages $\{m_1,m_2\}$ for any $m_1,m_2$ are sampled $\frac{16}{\underline{\Delta}^2}\log(NTM)$ times in some context is bounded by
\begin{equation}\nonumber
\begin{array}{ll}
\Pr\{\text{all pairs of messages are recommended to more than} \\
\text{~~~~~ $\frac{16}{\underline{\Delta}^2}\log(NTM)$ users in context $k$} \} \\
\geq 1 -  \frac{M^2}{(NMT)^4} \geq 1 - \frac{1}{(NT)^2} . 
\end{array}
\end{equation}
The probability of mis-clustering messages in content types can be bounded by
\begin{equation}
\begin{array}{ll}
\Pr\{\text{mis-classification at the end of phase 1} \} \\
\leq M \Pr\{|{\bar r}_t(k,m) - \mu(k,m)| \ge \frac{\underline{\Delta}}{4}\} + \frac{1}{(NT)^2}\\
\leq 2M \exp(-2 \frac{16}{\underline{\Delta}^2}\log(NTM)  \frac{\underline{\Delta}^2}{16})  + \frac{1}{(NT)^2} \\
\leq \frac{1}{(NT)^2} +  \frac{1}{(NT)^2}.
\end{array}
\end{equation}

Therefore, the regret caused by mis-classification is bounded by $NT( \frac{1}{(NT)^2} +  \frac{1}{(NT)^2}) \le 1$ and the regret caused by the exploration phase 1 is $\frac{32NKM}{\underline{\Delta}^2}\log(NTM)$. Combining these  with the performance of group recommendation in Section~\ref{sec:multi}, we get the result.